# An Insight into Spreadsheet User Behaviour through an Analysis of EuSpRIG Website Statistics

Grenville J. Croll
EuSpRIG – European Spreadsheet Risks Interest Group
grenvillecroll@gmail.com
v1.15

**ABSTRACT**

*The European Spreadsheet Risks Interest Group (EuSpRIG) has maintained a website almost since its inception in 2000. We present here longitudinal and cross-sectional statistics from the website log in order to shed some light upon end-user activity in the EuSpRIG domain.*

## 1 INTRODUCTION

EuSpRIG was founded in March 1999 as a collaboration between spreadsheet researchers at the University of Greenwich, the University of Wales Institute Cardiff and HM Customs & Excise. Its mission was to bring together academics, professional bodies and industry practitioners throughout Europe to address the ever-increasing problem of spreadsheet integrity.

EuSpRIG held its first formal meeting in the offices of Her Majesty's Customs and Excise as it was then known, in central London in January 2000. The group constitution was drawn up and the first officers appointed. The meeting decided to organise and hold a conference on Spreadsheet Risks at Greenwich University, London that July. There was some concern that anyone other than the organisers would turn up. The conference was, however, a great success with 50 delegates, some excellent papers and presentations, followed by a very enjoyable conference dinner.

EuSpRIG has by now held eleven conferences: Greenwich, London (2000, 2005, 2007, 2008, 2010); Amsterdam (2001), Cardiff (2002), Dublin (2003), Klagenfurt (2004), Cambridge (2006), Paris (2009).

EuSpRIG established a website in the year 2000, shortly after EuSpRIG's formal establishment [Chadwick, 2003]. The EuSpRIG brand identity was revised and went online in October 2009 at the same time as a major update to the website (www.eusprig.org ) which significantly extended the content. The input to this analysis is the complete web traffic data for calendar year 2010, with basic web site hits from 2004-2010 by month.

## 2. OBJECTIVES

There is very little work in the public domain outlining the characteristics and behaviour of spreadsheet users other than that which, by necessity, is part of spreadsheet error research [Panko, 2000] [Panko & Ordway, 2005] [Croll, 2005] [Caulkins et al, 2007]. An important exception is a large scale survey of MBA spreadsheet users [Baker et al, 2005] undertaken as part of the Spreadsheet Engineering Research Project (SERP). Their paper also briefly reviews spreadsheet user survey work over the past 20 or so years.

EuSpRIG objectives are clearly stated on their website:

> *EuSpRIG offers Students, Professors, Directors, Managers and Professionals in all disciplines the World's only independent, authoritative and comprehensive web based information on the current state of the art in Spreadsheet Risk Management.*
>
> *EuSpRIG is the largest source of information on practical methods for introducing into organisations processes and methods to inventory, test, correct, document, backup, archive, compare and control the legions of spreadsheets that support critical corporate infrastructure.*

This paper seeks to extract pertinent information from spreadsheet users internet searches which have landed on the EuSpRIG website in order to inform future research about end-user computing in the EuSpRIG domain. Clearly, many related internet searches will not land on a EuSpRIG web page. We use Google page ranking statistics to qualitatively infer the likelihood that particular internet searches will land on a EuSpRIG web page.

The analysis of website logs is a common activity in the pursuit of commercial profit. We assume that investigation of the EuSpRIG website logs provides useful information about spreadsheet users thoughts and actions.

## 3 SITE HITS SUMMARY

The first analysis of EuSpRIG web site traffic was reported to the EuSpRIG executive in June 2006. It reported site hits (i.e. unique visitors) by month from January 2004 (Figure 1). Recorded site hits at the beginning of the period in January 2004 were 478 per month, which grew over a period of 30 months to 1949 hits in June 2006. A logarithmic regression model proved a good fit (Adj. $R^2$=80%) showing a compound growth of 60% per annum over the 30 month period. Site traffic came from diverse sources and there were no particular outliers which had to be accommodated in this early longitudinal analysis. The hit counter was changed in July 06 with no material changes observed.

**Figure 1**

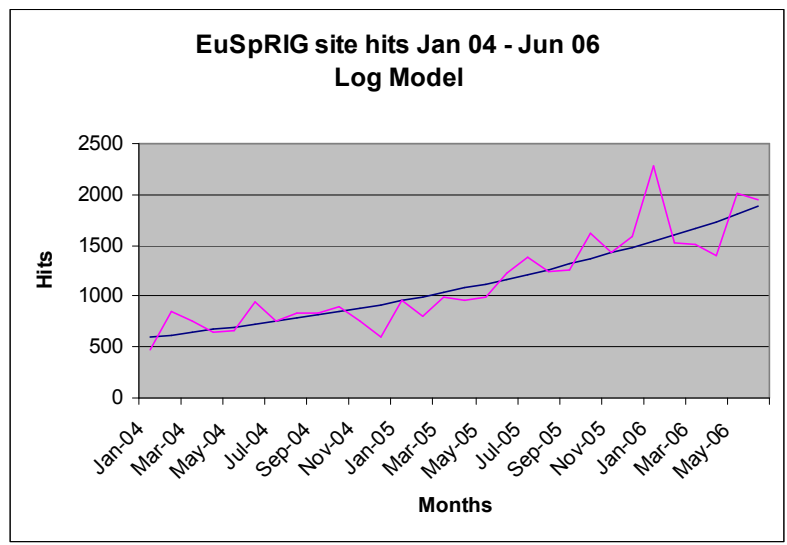

As is the case with exponential models, they soon come to an end. The second analysis of monthly website statistics, which is reported to the EuSpRIG executive in this report, occurred in January 2011. Exponential growth ceased in February 08, to be replaced after a short period of decline by slower probably linear growth and monthly web site traffic of 2750 hits per month. There were three apparent exceptions to this pattern.

**Figure 2**

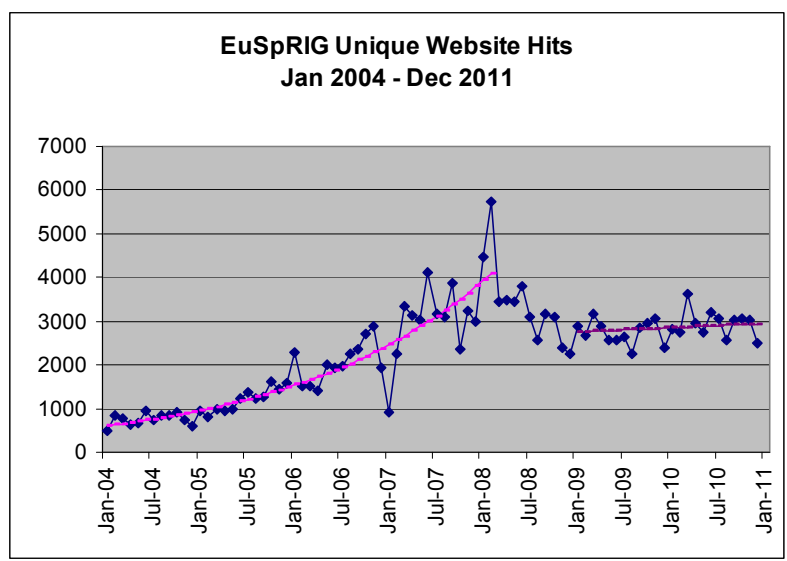

The first exception to the normal pattern of web site traffic occurred in February 2007, when there was a notable one month decline in traffic for which no explanation is offered. The second exception followed an article by Dominic Connor in www.theregister.co.uk in February 2008 which generated about 3000 hits [Connor, 2008]. The timing of this article corresponded to the onset of the Global Financial Crisis. The third exception occurred in the early days of February 2011, two months after the bulk of this analysis had been done, when a link to the EuSpRIG website was posted on Twitter www.twitter.com generating about 1,200 additional hits.

In Figure 2 we have extended the log model to include the period up to February 08. Annual compound growth was slightly in excess of 60% for this extended period and the model offered an improved fit (Adj. $R^2$=86%). We also show the slower linear growth from January 09.

The total number of unique visitors to the EuSpRIG website 2000-2010, including those visiting before the web counter was established, is approximately 250,000. Visitor numbers grew from 32,915 in 2009 to 35,327 in 2010, an increase of 7.3%.

**4 SITE HITS BY GEOGRAPHY**

Over a period of 12 months, the majority of countries are represented by at least one visitor to the EuSpRIG website as recorded by Google Analytics. By December 2010, of 13,795 total page downloads, 99.35% were accounted for by the following 25 countries:

**Table 1 EuSpRIG 2010 page downloads by domain/country**

| COUNTRY | DOMAIN | PAGES | % | %CUMUL |
|---|---|---|---|---|
| United States | us | 10776 | 78.12% | 78.12% |
| Australia | au | 805 | 5.84% | 83.95% |
| Unknown | ip | 729 | 5.28% | 89.24% |
| Great Britain | gb | 392 | 2.84% | 92.08% |
| European Union | eu | 263 | 1.91% | 93.98% |
| Germany | de | 166 | 1.20% | 95.19% |
| Netherlands | nl | 140 | 1.01% | 96.20% |
| Canada | ca | 66 | 0.48% | 96.68% |
| China | cn | 60 | 0.43% | 97.11% |
| Spain | es | 46 | 0.33% | 97.45% |
| Japan | jp | 38 | 0.28% | 97.72% |
| Hong Kong | hk | 31 | 0.22% | 97.95% |
| Belgium | be | 28 | 0.20% | 98.15% |
| Ireland | ie | 22 | 0.16% | 98.31% |
| France | fr | 21 | 0.15% | 98.46% |
| Singapore | sg | 20 | 0.14% | 98.61% |
| Switzerland | ch | 19 | 0.14% | 98.75% |
| South Korea | kr | 16 | 0.12% | 98.86% |
| Italy | it | 14 | 0.10% | 98.96% |
| Norway | no | 14 | 0.10% | 99.06% |
| Israel | il | 12 | 0.09% | 99.15% |
| Sweden | se | 10 | 0.07% | 99.22% |
| New Zealand | nz | 7 | 0.05% | 99.28% |
| South Africa | za | 6 | 0.04% | 99.32% |
| Ethiopia | et | 5 | 0.04% | 99.35% |

The United States, Canada, European Union, Australia and Japan account for the vast majority (93%) of page downloads. US dominance is almost overwhelming. European page downloads were 8.4%.

## 5 SEARCH PHRASES

### 5.1 SPECIFIC SEARCH TERMS

During the course of 2010, Google Analytics recorded that there were 12,520 unique search terms used which resulted in a visit, however brief, to the EuSpRIG website. The top 30 main search terms are given in Table 2 (note that minor spelling variations have been corrected & amalgamated for the more frequently occurring terms). If we regard the search terms of Table 2 as the "Specific" search terms, then these comprise 34.5% of the total. The "Generic" search terms, numbering approximately 6,500, comprise 65.5% of the total search terms.

In the final column of Table 2 we give the Google Page Rank of each search term. The Google Page Rank was obtained at the time of this analysis by searching again for each term in Column 1 of Table 2 and noting the Page Rank. So for example the Google Search term "EuSpRIG" gave www.eusprig.org as the first listed result. Likewise, the Google search term "importance of spreadsheet" gave www.eusprig.org as the 51[st] listed result. Clearly, the Google Page Rank will change over time and the manner and timing of such changes over a period may provide useful information into the state of art and the spreadsheet end-users mindset. There was little change in the page rankings of Table 2 in the approximate six month period between draft and final submission of this paper.

## Table 2 EuSpRIG 2010 Main Search Terms

| SEARCH TERM | HITS | % | % CUML | GOOGLE |
|---|---|---|---|---|
| eusprig | 1209 | 9.7% | 9.7% | 1 |
| spreadsheet modelling | 833 | 6.7% | 16.3% | 2 |
| spreadsheet errors | 417 | 3.3% | 19.6% | 1 |
| european spreadsheet risks interest group | 280 | 2.2% | 21.9% | 1 |
| importance of spreadsheet | 263 | 2.1% | 24.0% | 51 |
| spreadsheet risks | 235 | 1.9% | 25.9% | 1 |
| spreadsheet modelling best practice | 209 | 1.7% | 27.5% | 3 |
| spreadsheet best practice | 98 | 0.8% | 28.3% | 1 |
| news stories about | 83 | 0.7% | 29.0% | 3 |
| stories about | 72 | 0.6% | 29.5% | 63 |
| best practice | 65 | 0.5% | 30.1% | 477 |
| group with stories | 62 | 0.5% | 30.6% | >300 |
| medical spreadsheets | 61 | 0.5% | 31.0% | 16 |
| how do you know your spreadsheet is right | 49 | 0.4% | 31.4% | 1 |
| what is spreadsheet modelling | 42 | 0.3% | 31.8% | 4 |
| errors in spreadsheets | 41 | 0.3% | 32.1% | 1 |
| excel spreadsheet errors | 32 | 0.3% | 32.4% | 1 |
| spreadsheet | 31 | 0.2% | 32.6% | 116 |
| people behind spreadsheet | 28 | 0.2% | 32.8% | 79 |
| modelling best practices | 27 | 0.2% | 33.0% | 1 |
| spreadsheet design best practices | 25 | 0.2% | 33.2% | 5 |
| pediatric anesthesia worksheet | 23 | 0.2% | 33.4% | 12 |
| http //www.eusprig.org/smbp.pdf | 22 | 0.2% | 33.6% | 1 |
| horror stories | 21 | 0.2% | 33.8% | 15 |
| spreadsheet mistakes | 21 | 0.2% | 33.9% | 1 |
| how do you do a spreadsheet | 20 | 0.2% | 34.1% | 8 |
| table of contents practice | 19 | 0.2% | 34.2% | >300 |
| news stories | 18 | 0.1% | 34.4% | 28 |
| spreadsheet problems | 18 | 0.1% | 34.5% | 2 |

We believe it is highly likely that end users unfamiliar with EuSpRIG who see the EuSpRIG site as the first ranked site for their search term will click on the EuSpRIG site reference. Note that for 11 (36%) of the above 30 search terms, EuSpRIG is the top ranked site and for 18 search terms (60%), EuSpRIG appears on the first page of Google results.

Note that the word "Excel" appears only once in Table 2 and the words "Test" and "Document", not at all. Note that the words "modelling" and "modeling" appear at almost the exact same frequency and are listed together.

It is interesting that site visits due to use of the groups unique or abbreviated name comprise only 12% of site visits. It is of further interest that the number of visitors reaching the EuSpRIG site by use of "Spreadsheet Error", "Spreadsheet Risk" and related search terms comprise only about 6% of visitors. The EuSpRIG group has recently attempted to move its focus away from its initial interest in Spreadsheet Errors to a wider mandate suggested by the term "Spreadsheet Risk Management" (SRM). The evidence from end user internet searches suggests that end user interest in EuSpRIG is far more broadly based than even SRM. It is fascinating that three of the top 25 routes into the EuSpRIG website are from searches where the Google Page Rank is 300 or more.

It is encouraging to note that site visitors using "Spreadsheet Modelling", "Spreadsheet Best Practice" and related terms comprise about 10% of the popular specific search terms.

**5.2 GENERIC SEARCH TERMS**

By inspection, the Generic search terms are fairly interesting. First of all, they are extremely numerous, there being approximately 6,500 of them in 2010 alone. Secondly, they are extremely diverse. It will not be possible to give a full analysis of the Generic Search terms here, and a more rigorous taxonomic approach could be adopted.

To give an idea of the diversity of terms which have caused searchers to arrive at the EuSpRIG web site we give in Table 3 a count of the frequency of occurrence of search terms.

**Table 3**

| FREQ | NUMBER |
|---|---|
| >20 | 37 |
| >10 | 37 |
| >5 | 72 |
| 4 | 71 |
| 3 | 124 |
| 2 | 519 |
| 1 | 5773 |

Thus there are 37 search terms which occur with a frequency of more than 20 times. All of these are documented in Table 2. Astonishingly, nearly six thousand visits to the EuSpRIG website in 2010 were by people who invented their own unique search term. Given the limited space available here, we document in Figure 3, search terms which occurred with a frequency of 4 or more, classified by a generic type which we manually assigned.

**Figure 3 – Classification of Some Search Terms**

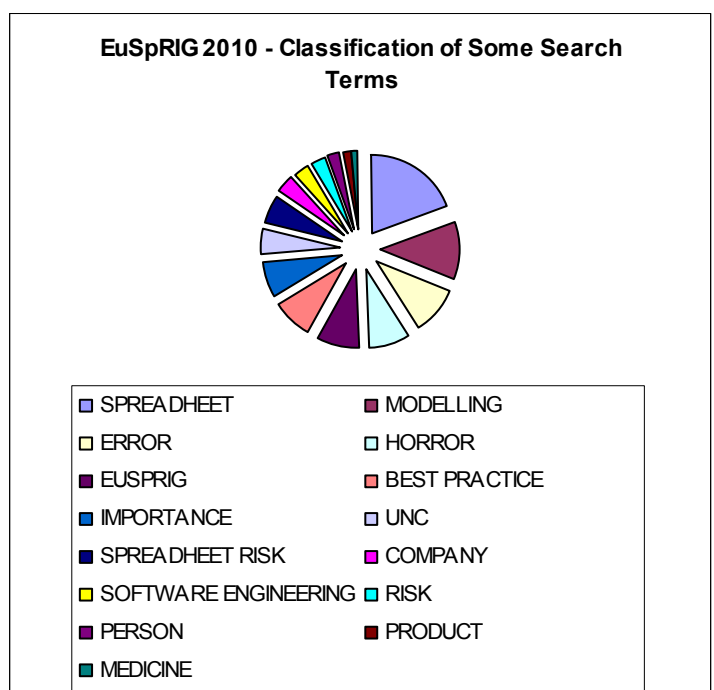

Clustering of search terms was certainly possible and included thematic variations and incorrect spellings (of which there were remarkably few).

Frequently occurring patterns within the generic search phrases include questions such as "How do I", "What is", "Why", "Who", "What to", together with a wide variation of every theme already reported and more. There are 700 single search terms that begin with the word "Spreadsheet". Table 4 gives the frequencies of occurrence of the questioning search terms.

**Table 4 EuSpRIG 2010 Questioning Search Terms**

| | | |
|---|---|---|
| How | 464 | |
| What | 349 | |
| Why | 86 | |
| When | 71 | |
| Who | 27 | |
| Which | 26 | |
| Where | 25 | |
| | | |
| Total Questions | 1048 | 8.3% |
| | | |
| Total Search Terms | 12520 | 100% |

## 5.3 GENERIC SEARCH PHRASE CURIOSITIES

There are a number of curious search terms which have lead end-users to the EuSpRIG website. One curiosity is where a user has typed a complete exam question into Google and then visited the EuSpRIG website. A good example is:

> "3. discuss an example of how you have used a spreadsheet to help you make a decision. discuss the spreadsheet logic the data inputs and calculated values that you used to make a decision.".

Another example is:

> "*in the plan section you are required to plan the layout of your spreadsheet. give your spreadsheet a title and using the information below start to think about what information is required. the layout of the spreadsheet should be clear and easy to read. remember to leave space for your bar graph. below is a list of the candidates who sat the exams and their results*."

Other curious examples are of the use of people or company names. These are substantially less frequent than one might have expected.

## 5.4 THEMATIC SEARCH TERMS

We used a simple spreadsheet to extract the frequency of occurrence of keywords relating to important themes within spreadsheet research from the 12,520 EuSpRIG 2010 search terms. The full software development lifecycle as it might be applied to spreadsheets has previously been established [Grossman & Ozluk, 2004].

### 5.4.1 Spreadsheet Testing

Search terms relating to testing spreadsheets [Pryor, 2004] [Panko, 2006] occurred 132 times – about 1% of the total. Table 5 gives the frequency of occurrence of the more frequently occurring terms relating to testing.

**Table 5- EuSpRIG 2010 Search Terms – "TEST"**

| | |
|---|---|
| Test Plan Spreadsheet | 14 |
| Spreadsheet Test Plan | 7 |
| Best Spreadsheet Test | 3 |
| Cross Foot Balance Test | 3 |
| Best practices spreadsheet testing | 2 |
| How to test a spreadsheet | 2 |
| Out of balance financial software testing | 2 |
| Spreadsheet practice test | 2 |
| Test plan for a spreadsheet | 2 |
| Test your spreadsheet | 2 |

The other 93 search terms relating to testing spreadsheets each occurred once. There were 47 occurrences of the use of the word "cross", as in cross testing, cross foot totals etc, a simple to implement but effective method of checking spreadsheets [O'Beirne, 2009].

5.4.2   Spreadsheet Documentation

There were just 48 occurrences of the word "Document" [Pryor, 2006] [Payette, 2006] in the 12,520 EuSpRIG 2010 search terms (0.38% ). All 48 search terms were different. A general lack of interest in documenting spreadsheets has been previously noted.

5.4.3   Spreadsheet Best Practice

It is pleasing to note that there were 756 occurrences (6%) of a search term that included the phrase "Best Practice" in the EuSpRIG 2010 search terms. 288 of these were unique. Table 6 lists the most frequently occurring "Best Practice" search terms.

**Table 6 – EuSpRIG 2010 Search Terms "BEST PRACTICE"**

| | |
|---|---|
| Spreadsheet modelling best practice | 153 |
| Spreadsheet best practice | 61 |
| Spreadsheet best practices | 31 |
| Modeling best practices | 27 |
| Spreadsheet design best practices | 25 |
| Best practice | 24 |
| Spreadsheet modeling best practice | 15 |
| End user computing best practices | 13 |
| Spreadsheet modeling best practices | 12 |
| Best practices | 11 |
| Best practice pdf | 10 |
| Best practice modelling | 9 |
| Modelling best practice | 9 |
| Best practice spreadsheets | 6 |
| Spreadsheet modeling for best practice | 6 |
| Spreadsheet modelling best practice ibm | 6 |
| Spreadsheets best practices | 6 |
| Best practices modeling | 5 |
| Best practice modeling | 4 |
| Best practice spreadsheet modelling standards | 4 |
| Spreadsheet modelling best practices | 4 |

Several of the above search terms are directed towards a popular Best Practice guide [Read & Batson, 1999] which is available on the EuSpRIG website.

5.4.4 Spreadsheet Engineering

There were 51 search terms out of 12,520 (0.4%) that contained the word "Engineer". Only some of them related to Spreadsheet Engineering [Grossman, 2002]. All 51 terms were different.

**6 SUMMARY AND CONCLUSION**

We have briefly introduced the European Spreadsheet Risks Interest Group (EuSpRIG) outlining its establishment and development over eleven years 2000-2010. We have analysed summary and detailed web data for the same period by quantity, geography and content. The content analysis was performed by looking at specific search terms, generic search terms, curiosities and important themes relating to the activities of EuSpRIG.

In 2010 there were about 35,000 unique visitors to the EuSpRIG website, a total which had been growing exponentially at 60% per annum in previous years, but is now growing more slowly at 7.3% per annum. This may reflect an initial interest followed by a maturation phase. It is curious, and perhaps merely a coincidence, that the period of decline in between the two differing growth cycles corresponds to the period of the recent global financial collapse [Croll, 2009]

Analysis of the 12,520 unique search terms which resulted in visitors accessing the EuSpRIG website in 2010 provides an interesting insight into the thoughts and interests of spreadsheet users worldwide.

Only a small proportion of these users – around 10% - search for important themes within the software engineering paradigm, namely testing, documentation and best practice. It is likely, but not certain (given EuSpRIG's generally high Google page ranking for these search terms), that these visitors to the EuSpRIG website are almost the only End-Users interested in these issues.

Based upon this analysis of the EuSpRIG 2010 web site statistics, it would appear to be the case that in 2010 only a few thousand people worldwide had a qualified interest in issues related to the integrity and quality of spreadsheets, best practices related to their development and the minimisation of risks related thereto. Given the ubiquity, importance and criticality of spreadsheets [Croll, 2005] within contemporary society, this is a continuing cause for concern which EuSpRIG quite rightly addresses.

**7 ACKNOWLEDGEMENTS**

The author thanks Patrick O'Beirne for his establishment of the original EuSpRIG web site, his colleagues on the EuSpRIG committee for their valuable contributions to the re-branding & re-writing of the EuSpRIG website and Tineke Warren of KINET e-solutions for its faultless reimplementation & maintenance. The author thanks the anonymous referees for their constructive comments.